\documentclass[aps, prl, showkeys, showpacs, reprint]{revtex4-1} 

\usepackage[usenames,dvipsnames]{color} 
\usepackage{amsmath} 
\usepackage{txfonts} 
\usepackage{upgreek} 
\usepackage{todonotes} 
\usepackage{amssymb} 
\usepackage{braket} 
\usepackage{array} 

\usepackage{enumitem}

\usepackage[unicode]{hyperref} 
\hypersetup{
    unicode=true,                                           
    a4paper=true,
    plainpages=false,
    pdffitwindow=true,                                      
    pdftitle={Enhanced optical cross section via collective coupling of atomic dipoles in a 2D array},                      
    pdfauthor={R J Bettles},                            
    pdfsubject={},                                          
    pdfkeywords={},
    colorlinks=true,                                        
    linkcolor=blue,                                         
    citecolor=blue,                                         
    filecolor=blue,                                         
    urlcolor=blue                                           
}
\usepackage{upgreek} 

\usepackage[caption=false]{subfig} 
\usepackage{graphicx} 
\usepackage[countmax]{subfloat}
\usepackage{import} 
\usepackage{transparent} 

\definecolor{MyGreen}{rgb}{0,0.5,0} 

\newcommand{\dvec}{\ensuremath{\mathrm{\mathbf{d}}}} 
\newcommand{\Evec}{\ensuremath{\mathrm{\mathbf{E}}}} 
\newcommand{\rvec}{\ensuremath{\mathrm{\mathbf{r}}}} 
\newcommand{\Rvec}{\ensuremath{\mathrm{\mathbf{R}}}} 
\newcommand{\bvec}[2]{\ensuremath{\boldsymbol{\mathrm{#1}}_{#2}}} 
\newcommand{\RL}{\ensuremath{R_{\mathrm{L}}}}  
\newcommand{\wL}{\ensuremath{w_{\mathrm{L}}}}  
\newcommand{\zR}{\ensuremath{z_{\mathrm{R}}}}  
\newcommand{\zL}{\ensuremath{z_{\mathrm{L}}}}  
\newcommand{\EL}{\ensuremath{E_{\mathrm{L}}}}

\graphicspath{ {../Figures/} }

\makeatletter
\newcommand{\colorcaption}[3][]{%
  \begingroup%
  \renewcommand{\@caption@fignum@sep}{ (color online). }%
  \caption[#1]{#2}
  \label{#3}
  \endgroup%
}
\makeatother

\begin{document}

\bibliographystyle{apsrev4-1}

\title{Enhanced optical cross section via collective coupling of atomic dipoles in a 2D array}
\author{Robert J. Bettles}\email{r.j.bettles@durham.ac.uk}
\author{Simon A. Gardiner}\email{s.a.gardiner@durham.ac.uk}
\author{Charles S. Adams}\email{c.s.adams@durham.ac.uk}
\affiliation{Joint Quantum Center (JQC) Durham--Newcastle, Department of Physics,  Durham University, South Road, Durham, DH1 3LE, United Kingdom}
\date{\today}

\begin{abstract}
Enhancing the optical cross section is an enticing goal in light-matter interactions, due to its fundamental role in quantum and non-linear optics. Here, we show how dipolar interactions can suppress off-axis scattering in a two-dimensional atomic array, leading to a subradiant collective mode where the optical cross section is enhanced by almost an order of magnitude. As a consequence, it is possible to attain an optical depth which implies high fidelity extinction, from a monolayer. Using realistic experimental parameters, we also model how lattice vacancies and the atomic trapping depth affect the transmission, concluding that such high extinction should be possible, using current experimental techniques.
\end{abstract}


\maketitle

Strong coupling between light and matter has been a long sought-after goal. Light-matter coupling can be conveniently characterized in terms of \textit{extinction\/} which corresponds to the probability for a medium to remove a photon from an incident field. For a single dipole, the highest recorded extinctions, of order $10\%$, have been achieved using individual molecules \cite{Wrigge2008} and atoms \cite{Tey2009} \footnote{Single dipole extinctions of $22\%$ were demonstrated for molecules by selecting only the coherent dipolar emission \cite{Wrigge2008}}
; single dipole extinction has also been demonstrated using ions \cite{Piro2011} and quantum dots \cite{Vamivakas2007}. The free-space extinction is typically limited by the focusing strength of a lens or mirror \cite{Tey2009}, and can be further enhanced using a waveguide or cavity thereby attaining the so-called \textit{strong coupling\/} regime associated with cavity QED \cite{Mucke2010,Thompson2013}. 
Replacing the single dipole with a high density ensemble of dipoles can have a dramatic effect on the optical response \cite{Dicke1954}. Coherent scattering between dipoles results in collective behavior, which can include enhanced or reduced scattering rates (superradiance or subradiance respectively) \cite{DeVoe1996,Scheibner2007,Rohlsberger2010,Goban2015a}, lineshifts \cite{Rohlsberger2010,Keaveney2012,Meir2014} and interference lineshapes \cite{Bettles2014a,Luk'yanchuk2010,Jenkins2013}. Recent experiments have shown that at high densities the dipole-dipole interaction in random atomic ensembles can significantly attenuate the optical extinction in both very hot ($\sim 100$ K) \cite{Keaveney2011} and cold \cite{Rath2010,Pellegrino2014a,Kemp} ($\sim 100$ $\upmu$K) atomic vapors.
Placing scatterers in a regular array formation can further enhance the cooperative response. Examples include near perfect extinction and transmission through arrays of gold nanorods \cite{Ghenuche2012}, linewidth narrowing in metamolecules \cite{Fedotov2010}, and extraordinary optical transmission (EOT) in hole arrays \cite{Ebbesen1998}. In addition to diffraction and interference effects, the coupling to collective and plasmonic modes plays a crucial role in explaining these phenomena \cite{Jenkins2012b,Martin-Moreno2001,GarciadeAbajo2007}. 
Cooperative broadening and shifts \cite{Bettles2014a,Kramer} as well as subwavelength excitation \cite{Jenkins2013} have been predicted in analogous atomic dipolar arrays, with the advantages that atomic systems allow easy access to the quantum regime, have much higher Q-factors, and significantly less non-radiative decay than the aforementioned plasmonic systems. 
In this Letter we show that atomic 2D arrays can also exhibit extreme variation in transmission depending on geometry. For certain \textit{magic lattice spacings}, high-fidelity extinction can occur, corresponding to an enhanced atom-light coupling which may open the door to exciting new applications in quantum simulation and information processing. 
Unlike the photonic bandgaps predicted in 3D atomic lattices \cite{Antezza2009,Antezza2013}, extinction in our system is due to a 
subradiant mode rather than a gap in the density of states.

Extinction, like many light-matter phenomena, is an interference effect. The total electric field at position $\rvec$, ${\Evec(\rvec) = \Evec_0(\rvec) + \sum_i \Evec_i (\rvec)}$, is the sum of the driving field, $\Evec_0(\rvec)$, and the fields radiated by the $N$ scatterers, $\sum_{i=1}^N \Evec_i (\rvec)$; \textit{extinction} of the driving field occurs when the driving and scattered fields interfere destructively. The scattered field from an electric dipole $\dvec_i$ located at $\rvec_i$ is $\Evec_i (\rvec) = \mathop{\mathsf{G}(\Rvec_i)} \dvec_i$, where $\mathop{\mathsf{G}(\Rvec_i)}$ is the dipole propagation tensor (Eq. (S1) in Supplemental Material \footnote{See Supplemental Material at [{URL}].}) and $\Rvec_i = \rvec - \rvec_i$. This dipole moment in turn is driven by the total local electric field, $\dvec_i = \alpha \Evec(\rvec_i)$, where $\alpha$ is the dipole polarizability. For a closed 2-level ${J=0\to J=1}$ atomic transition (e.g.\ Sr \cite{Olmos2013} or Yb \cite{Fukuhara2009a}), the polarizability takes the form $\alpha = -\alpha_0 /[(\Delta/\gamma_0)+\textrm{i}]$ where $\alpha_0=6\pi\varepsilon_0/k_0^3$, $\varepsilon_0$ is the permittivity of free space, $\lambda_0=2\pi/k_0$ is the wavelength of the dipole transition, $2\gamma_0$ is the dipolar scattering rate and $\Delta=\omega-\omega_0$ is the detuning of the driving frequency $\omega$ from the transition frequency $\omega_0$. A similar treatment can applied to plasmonic nano-resonators \cite{Jenkins2012a,Hopkins2013}. The linear response of $\dvec_i$ to $\Evec$ implies weak driving and means our model is closely equivalent to a set of damped driven classical oscillators \cite{Svidzinsky2010}. The weak driving limit can nonetheless be used to predict the extinction occurring in the quantum limit.
Optimizing this extinction involves matching the spatial \cite{Zumofen2008,Tey2009,Fischer2014} and temporal \cite{Aljunid2013} modes of the incident field to the field scattered by the dipoles. The scattered field of a single dipole has a very similar spatial mode profile to a Gaussian beam tightly focused on that dipole \cite{Note2}. The difference between the two fields along the axis of propagation of the Gaussian beam is simply a numerical factor $k^2 w_0^2/3$ where $k=2\pi/\lambda$ is the beam wavenumber (we assume the rotating wave approximation, and hence $k\simeq k_0$) and the beam waist $w_0$ is the $1/\textrm{e}$ radius at the focus. Maximizing the overlap would require a tightly focused beam \cite{Zumofen2008,Tey2009} (using e.g.\ a high numerical aperture lens) with waist of order $w_0 \simeq 0.3\lambda$ --- far beyond the reach of conventional free-space lenses. The alternative we propose in this Letter is to replace the single dipole with a monolayer of dipoles, which can exhibit near $100\%$ extinction without the need for such strong focusing. If combined with Rydberg blockade this could be employed to realize a high fidelity photonic gate \cite{Paredes-Barato2014}.

\begin{figure}[t]
\begin{center}
\includegraphics[trim = 0mm 0mm 0mm 0mm, clip, width=8.6cm]{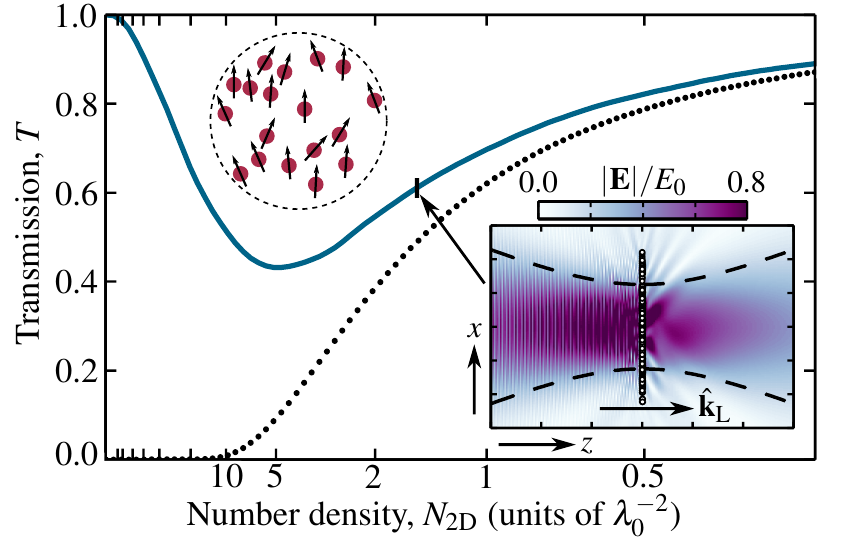}
\colorcaption{ 
Resonant optical transmission of a Gaussian beam through a random 2D monolayer of $N=100$ interacting dipoles. As the 2D number density $N_{\textrm{2D}}$ increases, the interacting monolayer (blue solid line) deviates from $T_{\mathrm{Ind}}$ (black dotted line), which assumes each dipole is a non-interacting opaque disk of cross sectional area $\sigma_0$. Each data point is averaged over 100 realizations. The beam waist is $w_0 \simeq 2.5\lambda$ and the collection lens has radius $R_{\textrm{L}} = 90\lambda_0$ and position $z_{\textrm{L}} = 150\lambda_0$.  (Inset) Weak cancellation of the total electric field magnitude $|\textbf{E}|$ in the $xz$ plane downstream of the monolayer ($N_{\textrm{2D}}\simeq 1.5\lambda_0^{-2}$). $x$ and $z$ vary between $\pm 6\lambda_0$ and $\pm 30\lambda_0$ respectively. The Gaussian beam propagates with vector $\hat{\textbf{k}}_{\textrm{L}} = \hat{\textbf{z}}$. The black dashed line shows the $1/\textrm{e}$ beamwidth and the white circles the atom positions. 
}{fig:1}
\end{center}
\end{figure}

The case of many dipoles is less trivial than for a single dipole, since now the local field experienced by each dipole is both the external driving field and also the fields scattered by the other $N-1$ dipoles, $\Evec (\rvec_i) = \Evec_0(\rvec_i) + \sum_{j\neq i} \Evec_j(\rvec_i)$. For an inhomogeneously broadened ensemble (e.g.\ a high-temperature thermal vapor \cite{Keaveney2012}), the sum of scattered fields $\sum_{j\neq i} \Evec_j(\rvec_i)$ can be replaced by an ensemble averaged mean field, resulting in, e.g., a geometry dependent \textit{cooperative Lamb shift} \cite{Friedberg1973,Keaveney2012,Javanainen2014}. The case we are interested in here is the homogeneously broadened regime (where atomic motion can be ignored \cite{Chomaz2012,Bettles2014a}), for which the recurrent scattering between dipoles must be included \cite{Javanainen2014}. Substituting $\dvec_i = \alpha \Evec(\rvec_i)$ into the equation for the local fields results in a set of coupled linear equations,
\begin{equation} \label{eq:dvec}
\dvec_i = \alpha \left( \Evec_0(\rvec_i) + \sum_{j\neq i} \mathop{\mathsf{G}(\Rvec_{ij})} \dvec_j \right),
\end{equation}
where $\Rvec_{ij} = \rvec_i-\rvec_j$.
These can be solved numerically for modest $N$ with arbitrary dipole positions and driving fields \cite{Ruostekoski1997a,Javanainen1999,Chomaz2012,Bettles2014a}.

To measure transmission and extinction, we calculate the total power passing through a lens downstream of the dipolar ensemble.
The power is related to the Poynting vector,
\begin{equation}
P = \frac{\varepsilon_0 c^2}{2} \int_{\textrm{L}} \Re [ \Evec \times \bvec{B}{}^* ] \cdot \textrm{d} \bvec{A}{},
\end{equation}
where $c$ is the speed of light, $\bvec{B}{} = \bvec{\hat{k}}{} \times \Evec/c$ is the B-field for an E-field with propagation unit-vector $\bvec{\hat{k}}{}$ and $\textrm{d}\bvec{A}{}=\mathop{\textrm{d}A }\hat{\textbf{z}}$ is the lens differential area element. We place the lens at $z_{\textrm{L}} = 150\lambda$ centered on $x,y = 0$. The lens radius $R_{\textrm{L}} = 90\lambda$ is large enough to avoid finite size effects \cite{Note2} whilst having a realistic numerical aperture ($\textrm{NA}=R_{\textrm{L}}/z_{\textrm{L}} = 0.6$). 
The driving field incident on the focusing lens has circular polarization vector $\hat{\bvec{\epsilon}{}}_+ = (\hat{x} + \textrm{i}\hat{y})/\sqrt{2}$. Strong focusing introduces small contributions from $\hat{\bvec{\epsilon}{}}_- = (\hat{x} - \textrm{i}\hat{y})/\sqrt{2}$ and $\hat{\bvec{\epsilon}{}}_z$, which we account for \cite{Note2,Tey2009}. The excited states $m_J=\{0,\pm 1\}$ are treated as degenerate, however driving a closed $m_J \leftrightarrow m_J+1$ transition gives quantitatively similar values for the optimal extinction.
We define transmission as the ratio of the power through the lens in the presence ($P$) and absence ($P_0$) of the dipoles, $T = P/P_0 = \textrm{e}^{- \sigma N_{\textrm{2D}}}$, where $\sigma$ is the extinction cross section and $N_{\textrm{2D}}$ is the 2D number density. Extinction is defined as $\epsilon \equiv 1-T$.
For low densities ($N_{\textrm{2D}} \ll \lambda_0^{-2}$, or $N_{\textrm{3D}} \ll \lambda_0^{-3}$ in 3D ensembles) the local field at each dipole is dominated by the external driving field since the scattered fields from neighboring dipoles in the far field decay with $1/(k_0 R_{ij})$, where $R_{ij} = |\Rvec_{ij}|$. In this case, the total extinction cross section is simply the cross section of an independent 2-level atom, ${\sigma_{\textrm{Ind}}=\sigma_0 /[1+(\Delta/\gamma_0)^2]}$, where $\sigma_0=3\lambda_0^2/(2\pi)$. As mentioned in the introduction, recent experiments in dense ($N_{\textrm{3D}} \gg \lambda_0^{-3}$) atomic vapors \cite{Keaveney2011,Rath2010,Pellegrino2014a,Kemp} have shown that dipole-dipole interactions reduce the cross section below the non-interacting value ($\sigma<\sigma_{\textrm{Ind}}$), increasing the transparency of the medium.

\begin{figure}[t]
\begin{center}
\includegraphics[trim = 0mm 0mm 0mm 0mm, clip, width=8.6cm]{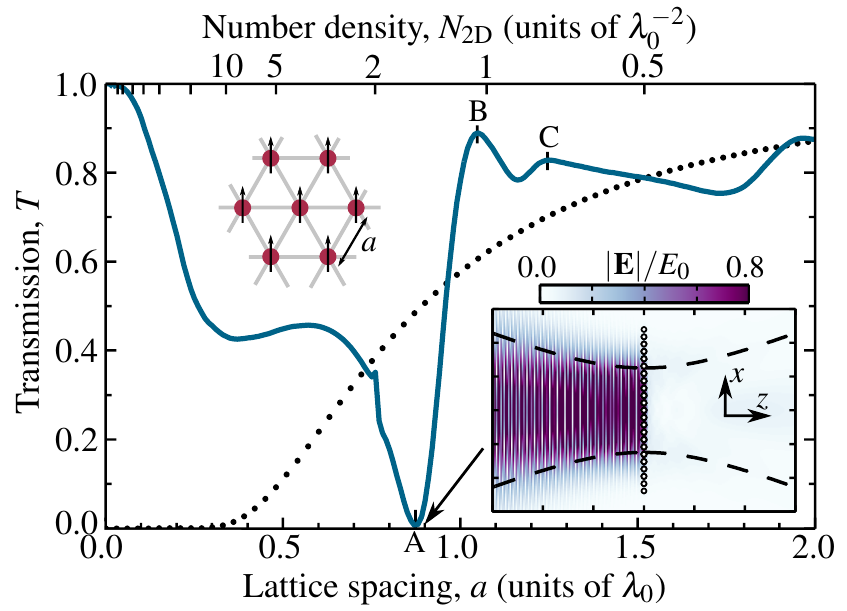}
\colorcaption{ 
Resonant optical transmission of a Gaussian beam through a triangular 2D array of $N=102$ interacting dipoles. 
Unlike the random monolayer in Fig.\ \ref{fig:1}, the transmission goes below and above $T_{\mathrm{Ind}}$ (black dotted line). A, B and C correspond to the lattice spacings used in Fig.\ \ref{fig:3}. (Inset) At $a=0.87\lambda_0$ ($N_{\textrm{2D}} \simeq 1.5\lambda_0^{-2}$) the dipole and driving fields almost perfectly cancel downstream of the lattice, resulting in less than $1\%$ transmission over the collection lens. The same parameters for the beam, lens and inset are used as in Fig.\ \ref{fig:1}.
}
{fig:2}
\end{center}
\end{figure}

As displayed in Fig.\ \ref{fig:1}, we start by considering resonant ($\Delta=0$) transmission through a 2D monolayer of uniformly randomly distributed atoms. The black dotted line plots the predicted transmission when ignoring dipole-dipole interactions, corresponding to the 2D limit of the familiar Beer-Lambert law $T_{\mathrm{Ind}}={\exp(-\sigma_{\textrm{Ind}} N_{\textrm{2D}})}$. In agreement with experiment \cite{Keaveney2011,Rath2010,Pellegrino2014a,Kemp}, the transmission increasingly deviates from the non-interacting Beer-Lambert value as the density increases. Shifts diverging as $1/R_{ij}^3$ between closely spaced dipoles result in a broadening and weakening of the overall cross section lineshape, reducing the resonant extinction (increasing transmission). 
It might therefore seem that interactions make the extinction worse. However, if we introduce spatial ordering to the atoms by confining them to a fixed regular (triangular) array, with one atom per site, we see in Fig.\ \ref{fig:2} that the transmission can be significantly lower than both the non-interacting and randomly distributed cases. Such an array could be realized in, e.g., an optical lattice in the Mott-insulator phase \cite{Greiner2002,Windpassinger2013} or spatial light modulator dipole trap array \cite{Nogrette2014}. For a particular magic lattice spacing ($a = 0.87 \lambda_0$) the extinction ($1-T$) is greater than $99\%$, corresponding to almost an order of magnitude increase in cross section ($\sigma\simeq 7\sigma_0$). 
Limits on the scattering cross section were discussed in \cite{Hugonin2015}.
The efficient cancellation of the electric fields downstream of the lattice can be seen in the inset of Fig.\ \ref{fig:2}, which is contrasted with the poorer extinction and significant scattering out of the beam in the random monolayer (inset, Fig.\ \ref{fig:1}).
The transmission minimum also corresponds to a reflection maximum, observable in the inset of Fig.\ \ref{fig:2}, as well as by calculating the power reflected back through the focusing lens at $z=-\zL$ (reflection $R\gg 98\%$). By slightly changing the lattice spacing ($a=0.87\lambda_0 \to 1.05\lambda_0$) the transmission increases from $<1\%$ to $\simeq90\%$. Consequently the monolayer can be switched between distinct transmission and reflection states, in the same spatial mode, which is the ideal starting point for a gate or all-optical transistor.

\begin{figure}
\begin{center}
\includegraphics[trim = 0mm 0mm 0mm 0mm, clip, width=8.6cm]{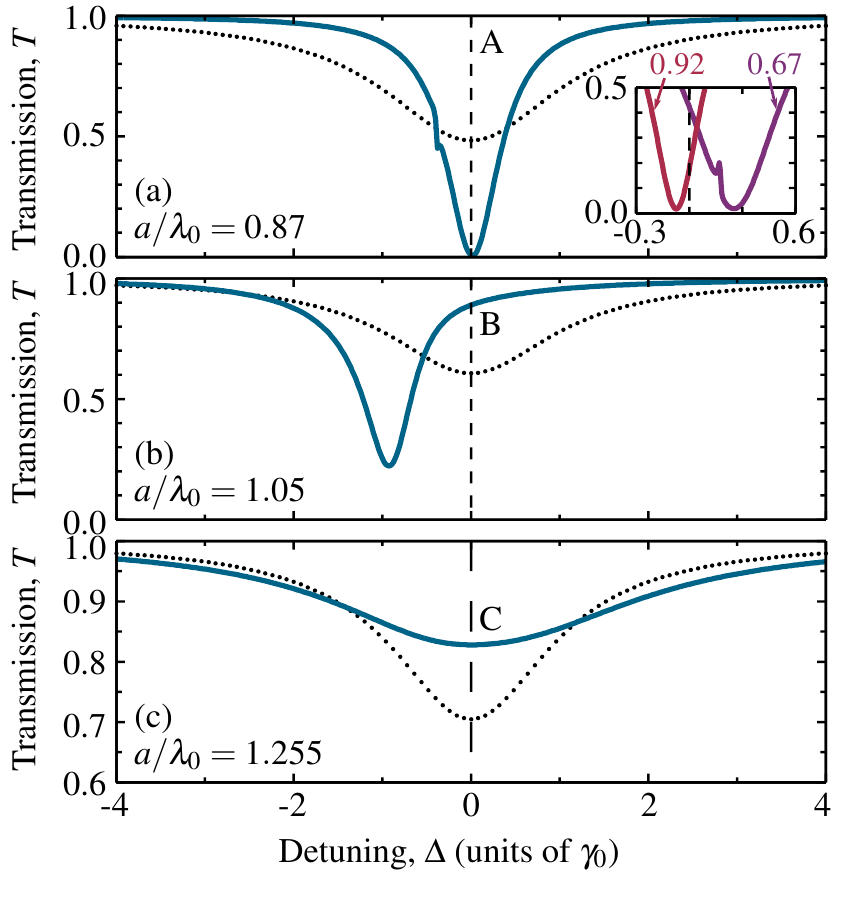}
\colorcaption{Transmission as a function of detuning through an $N=102$ triangular lattice of interacting dipoles. 
The lattice spacings in (a--c) correspond to those labeled A, B and C in Fig.~\ref{fig:2}. The blue solid lines plot the full interacting transmission through a lens of radius $\RL=90\lambda$ at position $\zL=150\lambda$. The black dotted lines show $T_{\mathrm{Ind}}$ (i.e., assuming no interactions). The vertical dashed lines at $\Delta=0$ have dash lengths $\Delta T =0.05$.
}
{fig:3}
\end{center}
\end{figure}

\begin{figure}
\begin{center}
\includegraphics[trim = 0mm 0mm 0mm 0mm, clip, width=8.6cm]{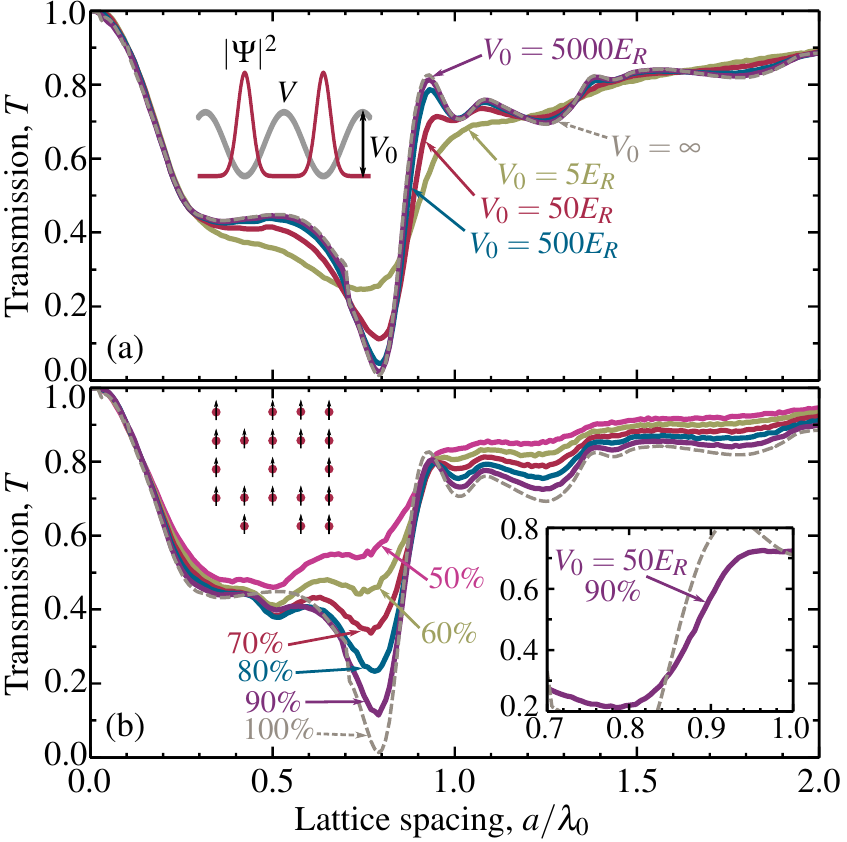}
\colorcaption{The effect of finite trap depth (a) and finite filling factors (b) on resonant optical transmission through a $10\times 10$ square lattice. 
(a) The trap depths are $V_0=\infty$ (grey dashed), $V_0=5000E_R$ (purple), $V_0=500E_R$ (blue), $V_0=50 E_R$ (red), and $V_0=5E_R$ (green), where $E_R$ is the recoil energy and the filling is $100\%$.
(b) The lattice sites are randomly occupied with filling factors of $100\%$ (grey dashed), $90\%$ (purple), $80\%$ (blue), $70\%$ (red), $60\%$ (green), and $50\%$ (pink), with $V_0=\infty$. The purple line in the inset is a combination of finite trap depth ($V_0=50E_R$) and $90\%$ filling.
 Each line is an average of several hundred realizations.
The same lens and beam parameters as in Fig.~\ref{fig:1} are used.
}
{fig:4}
\end{center}
\end{figure}

We now address why there is a magic spacing that produces optimal extinction.
In Fig.\ \ref{fig:3} we plot the transmission as a function of detuning at the points labeled A, B and C in Fig.\ \ref{fig:2}. 
The behavior of the interacting lineshapes (blue solid lines) is determined by the eigenmodes of Eq.\ (\ref{eq:dvec}). 
Each eigenmode contributes a shift $\Delta_l$ and linewidth $\gamma_l$ proportional to the real and imaginary parts of its eigenvalue, respectively \cite{Bettles2014a,Hopkins2013}.
The transmission behavior in Fig.\ \ref{fig:2} corresponds to the value of the transmission at $\Delta=0$, indicated by the vertical dashed lines in Fig.\ \ref{fig:3}.
In Fig.\ \ref{fig:3}(a) the lineshape is dominated by two nearly degenerate modes with halfwidths $\gamma_l=0.37\gamma_0$ centered at $\Delta_l\simeq 0$. 
Extinction cross section scales inversely with linewidth, so subradiance ($\gamma<\gamma_0$) results in an enhanced extinction.
This, combined with the maximal extinction at $\Delta=0$, results in the transmission minimum at $a=0.87\lambda_0$ (point A in Fig.\ \ref{fig:2}).
By changing the detuning of the driving field however, we can select a range of spacings over which large extinction is still possible [$\epsilon>98\%$ for $0.67<a/\lambda_0<0.92$, see inset in Fig.\ \ref{fig:3}(a)]. 
Figs.\ \ref{fig:3}(b) and (c) correspond to the local transmission maxima at points B and C in Fig.\ \ref{fig:2}. 
Whilst the peak extinction in (b) is still around $80\%$, it is shifted off resonance, so the extinction at $\Delta=0$ is small.
In (c) the lineshape is centered on $\Delta=0$, although it is now superradiant ($\gamma\simeq 2\gamma_0$) and so the peak extinction is reduced.

Large peak extinctions on resonance ($\Delta=0$) are also possible in square ($\epsilon>98\%$ at $a=0.79\lambda_0$, Fig.\ \ref{fig:4}) and hexagonal ($\epsilon>98\%$ at $a=0.6\lambda_0$) lattices with $N\sim 100$, providing further choice of trapping geometry. 
The complexity of the long range many body coupling responsible for this behavior means an analytic treatment is beyond the scope of this Letter.
We do however observe trends, for example the position of the magic lattice spacing increases with packing efficiency ($a/\lambda_0=\{0.6,0.79,0.87\}$ for hexagonal, square and triangular lattices respectively).

When considering a possible realization of this in an atomic experiment, it is necessary to consider how effects such as finite trapping depth [Fig.\ \ref{fig:4}(a)] and imperfect filling [Fig.\ \ref{fig:4}(b)] affect the extinction. We model finite trapping depth $V_0$ by treating each atomic wavefunction as a ground state harmonic oscillator \cite{Note2,Jenkins2012}. Averaging hundreds of realizations, atomic positions are sampled as Gaussian random variables centered on each lattice site with standard deviation related to $V_0$. Typical trap depths in Mott-Insulator experiments lie in the range $V_0=(20-50)E_R$ \cite{Bakr2009,Fukuhara2009a,Sherson2010,Bakr2011} ($E_R$ is the recoil energy \cite{Note2}), although $V_0\sim 10^3 E_R$ is possible \cite{Bakr2009,Sherson2010}. Filling efficiency greater than $90\%$ can be achieved \cite{Bakr2010,Sherson2010,Bakr2011,Lester2015a}, which when combined with a trap depth of $V_0=50E_R$ [Fig.\ \ref{fig:4}(b), inset], still gives a significant range in transmission [$(21\pm 5)\%$ to $(72 \pm 2)\%$ between $a\simeq 0.8\lambda_0$ and $a\simeq0.95\lambda_0$]. The extinction is also robust to small changes in the direction of incidence of the laser; rotating the incident laser $10^{\circ}$ from the normal of a $10\times 10$ square lattice still produces a peak extinction of over $90\%$.

The number of lattice sites does not have to be large to observe strong extinction; a $4\times 4$ perfect square lattice peaks at $\epsilon=96\%$ (for $w_0=\lambda$). 
With $100\%$ filling, increasing the atom number increases the peak extinction. The optimal beamwidth for maximizing the extinction scales with $\sqrt{N}$ ($w_0\simeq 2.5\lambda$ \footnote{The vector field propagation model results in a beam waist of $w_0\simeq 2.52 \lambda$ \cite{Note2}.} optimizes the extinction for square and triangular lattices with $N\simeq 100$). However, for $50\%$ filling as in Fig.\ \ref{fig:3}(b), adding more lattice sites (e.g.\ $200$ sites with $100$ vacancies) makes little difference to the transmission, meaning high filling factors are essential for high extinction.

In conclusion we have demonstrated numerically how the strong cooperative response of a 2D lattice of interacting dipoles can allow for very high extinctions (close to $100\%$) without the need for high densities, large atom numbers, or strong focusing. The cavity-like dependence on spacing between atoms in these periodic lattices results in a strong dependence on the lattice spacing. Thanks to its efficient packing the triangular lattice performs best, with a highly tunable transmission of between  $<1\%$ and $90\%$ for a small change in lattice spacing.
This work demonstrates further that the presence of interactions significantly modifies the optical response of a medium. Building on previous works in random gases \cite{Keaveney2011,Chomaz2012,Pellegrino2014a,Kemp}, we show that adding structure to the atom positions can significantly enhance such effects. By combining with Rydberg blockade one could realize a dipolar QED (dQED) analogue of the strong coupling regime in cavity QED, with potential applications for quantum non-linear optics.

\begin{acknowledgements}
We thank M.\ Greiner, C.\ Genes, H.\ Ritsch, S.\ Kr\"{a}mer, J.\ Ruostekoski and R.\ Kaiser for helpful discussions. 
We acknowledge funding from the UK EPSRC (Grant No.\ EP/L023024/1). The data presented  in  this  paper  are  available  at \doi{10.15128/vt150j378}. 
\end{acknowledgements}


%


\pagebreak
\clearpage

\bibliographystyle{apsrev4-1}

\widetext
\begin{center}
\textbf{\large Supplemental Material for \\ Enhanced optical cross section via collective coupling of atomic dipoles in a 2D array}
\end{center}
\setcounter{equation}{0}
\setcounter{figure}{0}
\setcounter{table}{0}
\setcounter{page}{1}
\makeatletter
\renewcommand{\theequation}{S\arabic{equation}}
\renewcommand{\thefigure}{S\arabic{figure}}
\renewcommand{\bibnumfmt}[1]{[S#1]}
\renewcommand{\citenumfont}[1]{S#1}

\section{Dipole field} 
The field at position $\rvec =  \rvec_i+\Rvec_i$ radiated from a dipole $\dvec_i$ at position $\rvec_i$ has the form \cite{Jackson1963}
\begin{eqnarray} \label{eq:EdFull}
\Evec_i (\rvec) &=& \mathsf{G}(\Rvec_i)\,\dvec_i =
 \frac{k^3}{4\pi\varepsilon_0} \mathrm{e}^{\mathrm{i}kR_i} \Bigg\{   (\hat{\Rvec_i}\times \dvec_i) \times \hat{\Rvec_i} \frac{1}{kR_i} 
 + [3 \hat{\Rvec_i} ( \hat{\Rvec_i} \cdot \dvec_i) - \dvec_i] 
\Bigg[\frac{1}{(kR_i)^3} - \frac{\mathrm{i}}{(kR_i)^2}  \Bigg]
\Bigg\}  ,
\end{eqnarray}
where $\varepsilon_0$ is the permittivity of free space and $k$ is the wavenumber of the radiated light. This form of the dipole field is used throughout the main text.

\section{Gaussian driving field}

In the paraxial approximation ($\rho\ll z$, where $\rho^2=x^2+y^2$), a Gaussian beam propagating along $z$ has the form
\begin{equation}\label{eq:EgParaxial}
\Evec_0(\rvec) = E_0 \frac{w_0}{w} \, \mathrm{e}^{\mathrm{i}[k(z+\rho^2/2R)-\zeta(z)]} \, \mathrm{e}^{-\rho^2/w^2} \hat{\bvec{\epsilon}{}},
\end{equation} 
where $w = w_0 \sqrt{1+z^2/z_{\mathrm{R}}^2}$ and $w_0$ are the 1/e beam radius at $z$ and $(z=0)$ respectively, $R=z+z_{\mathrm{R}}^2/z$ is the beam curvature, ${\zeta(z)=\arctan (z/z_{\mathrm{R}})}$ is the Gouy phase and $\zR=\pi w_0^2/\lambda$ is the Rayleigh range. However, the choice of focusing parameters in this Letter ($w_0=2.5\lambda$, $\zL=150\lambda$) means we are not fully in the paraxial limit and instead need to model the full vector field propagation. 

The following treatment follows closely that in \cite{Tey2009,VanEnk2001}, in which more details can be found. We start with a laser beam incident on a focusing lens a distance $\zL$ downstream of the atomic plane. 
The beam has electric field profile ${\bvec{E}{\mathrm{in}} = \mathrm{i}\EL \mathop{\mathrm{e}^{-\rho^2/\wL^2}} \hat{\bvec{\epsilon}{}}_+}$, where $\wL$ is the beam radius at the lens and ${ \hat{\bvec{\epsilon}{}}_+ = (\hat{\mathbf{x}}+\mathrm{i}\hat{\mathbf{y}})/\sqrt{2}}$ is a unit circular polarization vector. 
The factor of $\mathrm{i}$ is included so that the field in the focus will be approximately real. 
As the field propagates through the lens, it acquires a phase and the wavevector $\hat{\bvec{k}{}}$ changes direction. 
The change in $\hat{\bvec{k}{}}$ introduces small contributions from polarizations ${ \hat{\bvec{\epsilon}{}}_- = (\hat{\mathbf{x}}-\mathrm{i}\hat{\mathbf{y}})/\sqrt{2}}$ and $\hat{\bvec{z}{}}$. The total field immediately after the lens is then
\begin{equation} \label{eq:Ein}
\Evec (\rho, \phi, z=- f) = \frac{E_{\mathrm{L}} \mathop{\mathrm{e}^{-\rho^2/w_{\mathrm{L}}^2}} }{\sqrt{|\cos\theta|}} \left( \frac{1 + \cos \theta}{2} \hat{\bvec{\epsilon}{}}_+ + \frac{\sin\theta}{\sqrt{2}} \mathrm{e}^{\mathrm{i}\phi} \hat{\bvec{z}{}} + \frac{ \cos \theta -1}{2} \mathrm{e}^{2\mathrm{i}\phi} \hat{\bvec{\epsilon}{}}_- \right)  
\mathop{\mathrm{\mathrm{exp}}} \left[-\mathrm{i} \left( k \sqrt{\rho^2 + f^2} - \pi/2 \right) \right],
\end{equation}
where $\rho^2 = x^2 + y^2$, $\phi = \tan^{-1}(y/x)$ and $\theta=\tan^{-1}(\rho/f)$ is the angle between the $-z$ axis and a point on the lens. The total field can therefore be decomposed into an orthogonal set of modes, $\Evec = \sum_{\mu} \kappa_{\mu} \bvec{E}{\mu}$, where $\mu = (k_t,s,m)$, $k_t=\sqrt{k^2-k_z^2}$ is the transverse wavevector component, $s=\pm1$ is the helicity and $m$ is an angular momentum index. This decomposition now allows us to propagate this field to any point behind the lens. The expansion coefficients $\kappa_{\mu}$ are 
\begin{eqnarray} \label{eq:kappa_mu}
\kappa_{\mu} = \delta_{m1} \pi k_t \int_0^{\infty} \mathrm{d}\rho_{\textrm{L}}\, \rho_{\textrm{L}} \frac{1}{\sqrt{\cos\theta_{\textrm{L}}}} \Bigg\{ &&\frac{sk+k_z}{k} \left( \frac{1+\cos\theta_{\textrm{L}}}{2} \right) J_0(k_t \,\rho_{\textrm{L}}) + \mathrm{i}\frac{\sqrt{2}k_t}{k} \left( \frac{\sin\theta_{\textrm{L}}}{\sqrt{2}} \right) J_1(k_t\,\rho_{\textrm{L}}) \nonumber \\ 
&& + \frac{sk-k_z}{k} \left( \frac{\cos\theta_{\textrm{L}}-1}{2} \right) J_2(k_t\,\rho_{\textrm{L}}) \Bigg\} \mathrm{exp} \left[ -\mathrm{i} \left( k\sqrt{\rho^2_{\textrm{L}} + f^2} - \pi/2 \right) - \frac{\rho^2_{\textrm{L}}}{\wL^2} \right],
\end{eqnarray}
where $J_m$ is the $m$th order Bessel function, $\rho_{\textrm{L}}$ is the radial position across the lens and $\theta_{\textrm{L}}=\tan^{-1}(\rho_{\textrm{L}}/f)$. The field components in the $\pm$ and $z$ polarizations are then
\begin{eqnarray}
E_+ (\rho,\phi,z) &=& \EL \sum_{s=\pm1} \int^k_0 \mathrm{d}k_t \frac{1}{4\pi} \frac{s k + k_z}{k} J_0(k_t\rho) \mathop{\mathrm{e}^{\mathrm{i}k_z (z+f)}} \kappa_{\mu} , \nonumber \\
E_z (\rho,\phi,z) &=& \EL \sum_{s=\pm1} \int^k_0 \mathrm{d}k_t (-\mathrm{i}) \frac{\sqrt{2}}{4\pi} \frac{k_t}{k} J_1(k_t\rho) \mathop{\mathrm{e}^{\mathrm{i}k_z (z+f)}} \mathop{\mathrm{e}^{\mathrm{i}\phi}} \kappa_{\mu} , \nonumber \\
E_- (\rho,\phi,z) &=& \EL \sum_{s=\pm1} \int^k_0 \mathrm{d}k_t \frac{1}{4\pi} \frac{s k - k_z}{k} J_2(k_t\rho) \mathop{\mathrm{e}^{\mathrm{i}k_z (z+f)}} \mathop{\mathrm{e}^{2\mathrm{i}\phi}} \kappa_{\mu} .
\end{eqnarray}
Using this method we calculate the electric field, ${\Evec = E_+ \hat{\bvec{\epsilon}{}}_+ + E_- \hat{\bvec{\epsilon}{}}_- + E_z \hat{\bvec{\epsilon}{}}_z}$, in the plane of the atoms as well as across the output collection lens.  

In order to obtain a beam waist at the focus of $w_0\simeq2.5\lambda$, we use the paraxial equation for beam radius ${\wL = w_0(1+(\zL\lambda/\pi\wL^2)^2)^{0.5}}$ to estimate the input beam radius required, $\wL= 19.26\lambda$. For this focusing strength $u=\wL/f=19.26/150\simeq0.13$, the authors in \cite{Tey2009} calculate that there should be a noticeable difference between the paraxial and full vector field profiles in the focal plane. This correction however is still small, with the additional polarization contributions being $|E_z|<0.04E_0$ and $|E_-|<0.002E_0$ and the waist of $|E_+|$ being $w_0\simeq 2.52\lambda$. The correction to the corresponding transmission calculations is of the order of a few percent. Because this difference is small, the E-field colormap insets in Figs. 1 and 2 were produced using the analytic paraxial field Eq.\ (\ref{eq:EgParaxial}) as this was significantly less computationally intensive whilst still demonstrating the important results. 

The integral in Eq.\ (\ref{eq:kappa_mu}) assumes a lens with infinite radius, although in practice for our choice of parameters the interval converges sufficiently ($0.01\%$) by $\rho_{\textrm{L}}^{\mathrm{max}} \simeq 50\lambda$ and so to a good approximation we can assume that the input lens has the same radius as the output collection lens ($\RL=90\lambda$).

\section{Extinction}

The origin of the extinction of the Gaussian driving field by the field from a single dipole can be seen by comparing the fields in Eqs.\ (\ref{eq:EdFull}) and (\ref{eq:EgParaxial}) along the $z$ axis. In the far field ($|z|\gg\{\lambda_0,\zR\}$), the total field has the form
\begin{equation} \label{eq:extinction}
\Evec (z) = \Evec_i (z) + \Evec_0 (z)  \simeq \mathrm{i} \frac{3 E_0 \mathrm{e}^{\mathrm{i}k z}} {2kz} \left[ 1 - \textrm{sgn}(z) \frac{k^2 w_0^2}{3}  \right] \hat{\bvec{\epsilon}{}},
\end{equation}
where $\textrm{sgn}(z)$ is the sign of $z$. The only difference between the two fields is a numerical factor, $k^2w_0^2/3$, indicating that in the condition when $w_0\simeq 0.3\lambda$, the two fields will efficiently cancel in the $+z$ direction, resulting in extinction. Whilst the full, non-paraxial solution over a finite lens radius including many atoms is much more complex, the simple paraxial model gives a useful picture of how mode matching between the fields results in extinction.

\section{Optical Lattice Trapping Depth}
Following the approach in \cite{Jenkins2012}, we assume the trapping potential confining the atoms in a square lattice has the form 
\begin{equation}
V = V_0 \left[ \sin^2 \left(\frac{\pi x}{a}\right) + \sin^2 \left(\frac{\pi y}{a}\right) \right],
\end{equation}
where ${V_0= s E_R}$ is the amplitude of the trapping potential, $E_R=\pi^2 \hbar^2/2ma^2$ is the lattice recoil energy and $m$ is the mass of the atom. We assume an infinitely deep confining trap in the $z=0$ plane. The atom on each lattice site occupies the ground state of the harmonic oscillator 
\begin{equation}
\Psi_i(\Rvec_i) = \frac{1}{(\pi^3 l^4 l_z^2)^{1/4}} \exp \left( -\frac{X_i^2 + Y_i^2}{2 l^2} - \frac{Z_i^2}{2l_z^2} \right),
\end{equation} 
where $\Rvec_i = (X_i,Y_i,Z_i)$ is the separation of atom $i$ from the $i$th lattice site, $l=a s^{-1/4}/\pi$ and $l_z=\sqrt{\hbar/m\/\omega_z}$. The atomic positions are sampled at random using the probability distribution $\rho_i(\rvec) = |\Psi_i(\Rvec_i)|^2$ which is a Gaussian with $1/\textrm{e}$ radius $l$ in the $xy$ plane.


%

\end{document}